\magnification=1200
\baselineskip=18truept
\input epsf

\def\preprint{Y}
\def\draftversion{N}
\def\cap{\hsize=4.5in}

\if \draftversion Y


\fi

\def\figure#1#2#3{\vskip .2in
\if \preprint Y \midinsert \epsfxsize=#3truein
\centerline{\epsffile{figure_#1_eps}} \halign{##\hfill\quad
&\vtop{\parindent=0pt \hsize=5.5in \strut## \strut}\cr {\bf Figure
#1}&#2 \cr} \endinsert \fi}

\def\figureb#1#2{\if \preprint N \midinsert \epsfxsize=#2truein
\centerline{\epsffile{figure_#1_eps}} \halign{##\hfill\quad
&\vtop{\parindent=0pt \hsize=5.5in \strut## \strut}\cr \cr \cr
\cr \cr \cr  {\bf Figure #1} \cr} \endinsert \fi}

\def\captionfour{\cap $F_N$ 
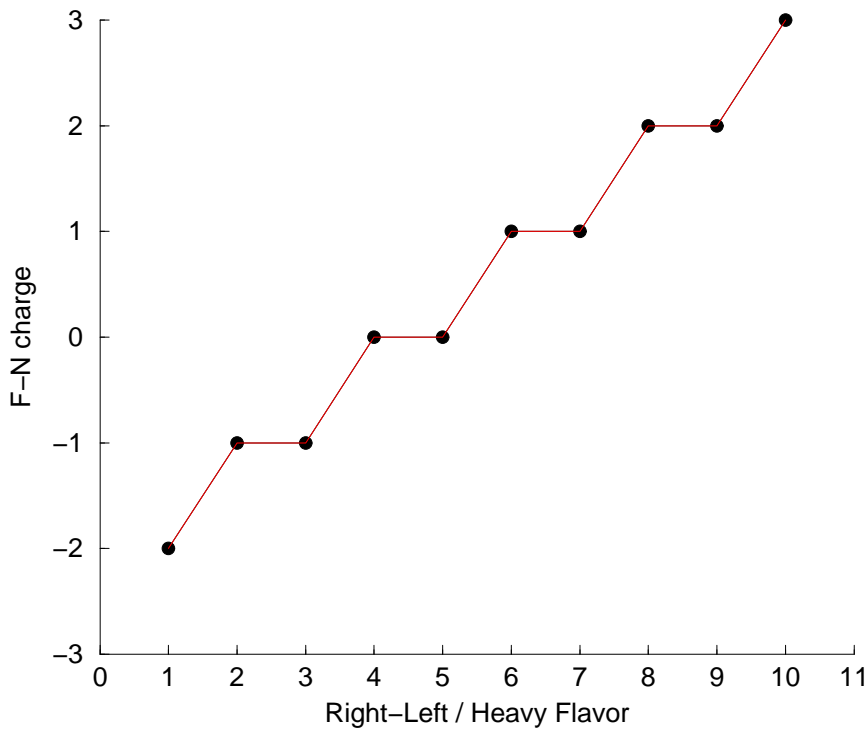
Charge assignments of the $\Phi_s$ fields for $s=1,2,...,2k$. Connecting
lines are there just to guide the eye.}
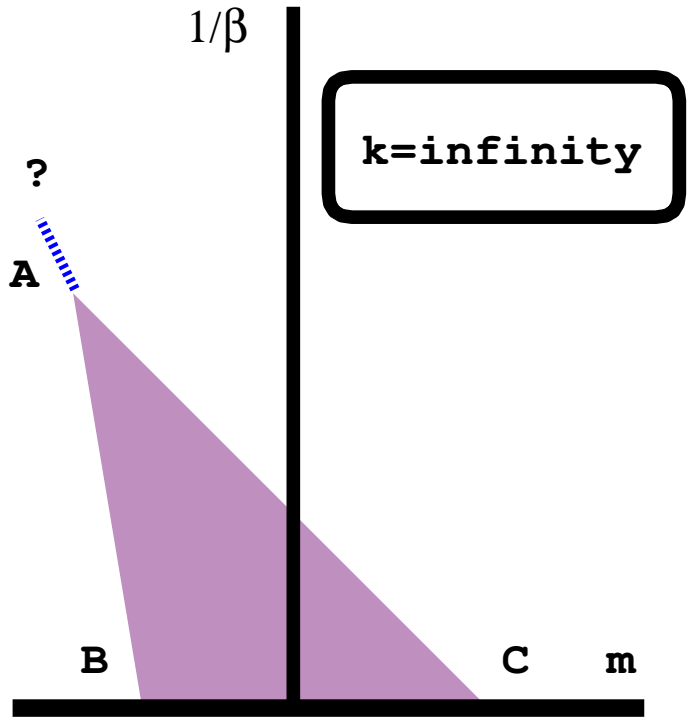
\def\captionone{\cap Schematic phase diagram for the overlap. In the
triangular area we have exact chiral symmetries. The possibility
for a continuing line from the point A is indicated. Along that line
one of the mesons could be massless, but 
exact chiral symmetry is not necessarily 
restored there. The phase diagram could be 
much more complicated. In the continuum limit one would approach
the interior of the segment BC for strictly massless quarks. 
To describe a continuum theory with a positive massive quark the endpoint
C should be approached from the outside. 
Immediately to the left of line AB  $d$ doublers become
massless, while the state associated with the origin
of momentum space becomes heavy. }
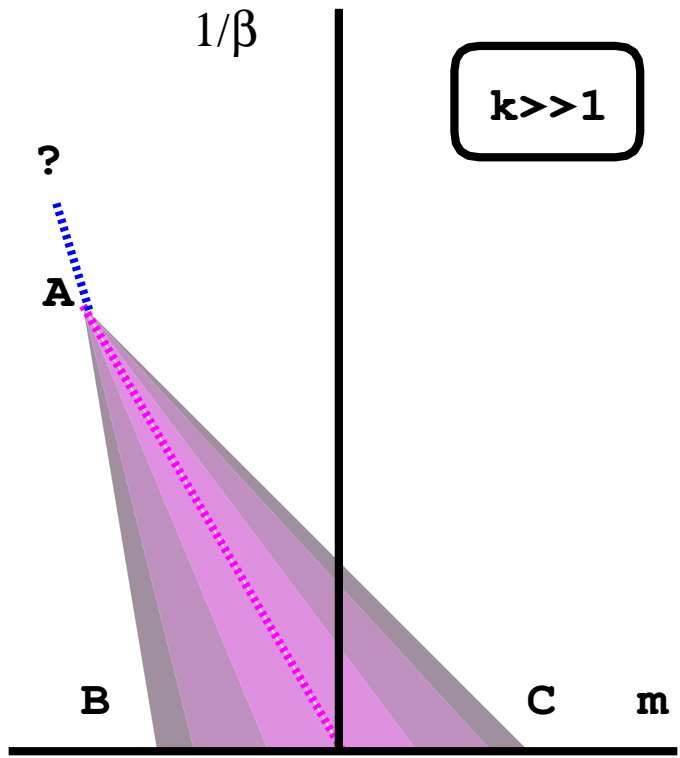
\def\captiontwo{\cap Schematic phase diagram for the truncated overlap,
with a large number of heavy flavors. In the
triangular area we have approximate chiral symmetries, with
the approximation improving towards the center. This is indicated
by the gradual change of shades. To the left of the line
AB one also expects approximate chiral symmetries but
they are accompanied by an unwanted increase in the number
of Dirac copies. 
On the center line
one of the ``pion'' states becomes massless, but,  exact
chiral symmetries exist only at the origin.}
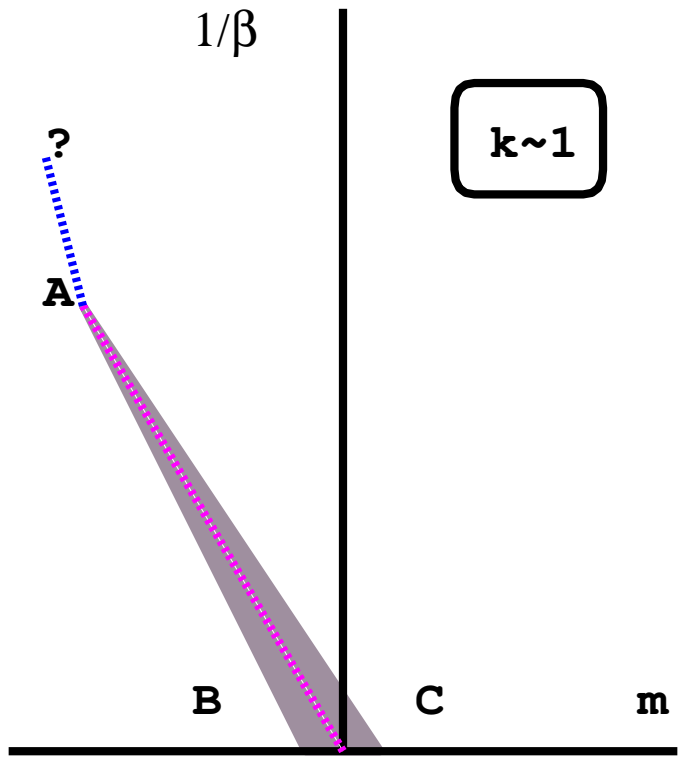
\def\captionthree{\cap Schematic phase diagram for the truncated
overlap with a small number of heavy flavors. The 
triangular area where we have approximate chiral symmetries has
shrunk relatively to figure 2. The mass of the quarks
decreases towards the center, but is much higher than the
mass in figure 2.  On the center line
one of the ``pion'' states becomes massless, but exact
chiral symmetries exist only at the origin. The shrinkage of the
triangle indicates the developing need for more and more accurate
mass tuning as the number of heavy flavors is decreased and the
regular Wilson case is approached. }

\def\recent{1}
\def\plbfirst{2}
\def\npold{3}
\def\npblong{4}
\def\blumsoni{5}
\def\jaster{6}
\def\vranas{7}
\def\kaplan{8}
\def\shamir{9}
\def\fursham{10}
\def\nninfmass{11}
\def\kiknn{12}
\def\narvra{13}
\def\boya{14}
\def\creutzbnd{15}
\def\shotkey{16}
\def\massineq{17}
\def\rebbi{18}
\def\ginspwil{19}
\def\curciv{20}
\def\aokizen{21}
\def\huetnn{22}
\def\hsu{23}
\def\odddim{24}
\def\montvay{25}
\def\nishimura{26}
\def\italiansusy{27}
\def\eyalgomez{28}
\def\seesaw{29}
\def\fn{30}
\def\smitswift{31}
\def\jansen{32}
\def\kogsinc{33}
\def\taniguchi{34}
\def\balaban{35}
\def\gibbs{36}

\line{\hfill RU-97-84}
\vskip 2truecm
\centerline{\bf Vector like gauge 
theories with almost massless fermions on the lattice.}
\vskip 1truecm
\centerline{Herbert Neuberger}
\vskip .5truecm

\centerline {Department of Physics and Astronomy}
\centerline {Rutgers University, Piscataway, NJ 08855-0849}
\vskip 1.5truecm

\centerline{\bf Abstract}
\vskip 0.75truecm

A truncation of the overlap (domain wall fermions) 
is studied and a criterion
for reliability of the approximation is obtained by
comparison to the exact overlap formula describing massless quarks.
We also present a truncated version of regularized,
pure gauge, supersymmetric models.
The mechanism for generating almost masslessness is shown to
be a generalized see-saw which can also be viewed as a version
of Froggatt-Nielsen's method for obtaining natural large mass
hierarchies. Viewed in this way the  mechanism
preserving the mass hierarchy naturally avoids preserving even approximately
axial $U(1)$. The new insights into the source of the mass hierarchy
suggest ways to increase the efficiency of numerical simulations
of QCD employing the truncated overlap. 
\vfill
\eject
{\centerline {\bf 1. Introduction}}

For massless quarks, QCD would have exact chiral symmetries. This
simple observation explains a large body of observations. 
At present we try
to solve QCD by numerical methods using a lattice regularization.
In the standard approach exact chiral symmetries cannot be
preserved by this regularization. 
This is disappointing but not debilitating since chiral symmetries
can be restored in the continuum limit. 
Still, there have been many attempts to get exact chiral symmetries
on the lattice, even just as a matter of principle. These attempts
often led to controversies and the issue is subtle. Recent progress
seems to have been achieved, and, although controversy still
exists on related issues, my impression is that most workers
would agree that there does exist a well defined, albeit
non-standard, way to preserve
global chiral symmetries exactly on the lattice. 
The price is quite high though:
The lattice model, 
although well defined, must be interpreted as containing an
infinite number of fermions
fields. Moreover, the expressions are complicated
and usage in practice appeared, only a year or two ago, quite
unlikely. However, recent developments raise the hope that 
this could change.

In a recent publication [\recent] a substantially
simpler formula for the effective
action in lattice vector-like theories with exact global chiral
symmetries was derived. It is based on the overlap which was 
developed [\plbfirst,\npold,\npblong] as a 
method to regulate chiral gauge theories
on the lattice. Obviously, as such, the 
overlap must contain the vector-like  as a particular case
case, where the chiral symmetries are not gauged.
Section 9 of [\npblong] contains a specific discussion of the properties
of the overlap in the vector-like context. What is 
new in [\recent], is that the expressions in [4] can be simplified.

Let us first briefly review the basic features of the overlap
relevant to the present context.
(Although the focus is on four dimensions we shall
try subsequently to write most equations in an arbitrary
even dimension $d$.) Formally, in the continuum path integral,
because the fermions enter
only bilinearly in the action, one can write any correlation function
as an average over gauge field configurations of an object
obtained by integrating first over the fermions. This Grassmann
integration produces a result factorized into two 
types of terms: A determinant (the exponentiated sum of vacuum
fermion diagrams) and a combination of entries
of inverse Dirac operators (propagators). 
In the regularized overlap the information contained in fermion 
vacuum diagrams is
stored in two (or one, see later) ground states of two auxiliary
quantum mechanics problems, parametrically dependent on the gauge
fields. The propagators are obtained by matrix elements 
of certain fermionic creation/annihilation operators
between the two ground states.
The crucial point is that each chiral component of a physical
field is represented by a separate set of such operators. 
This implies immediately
exact chiral symmetries, as the factorization is exact
and the system acts as if it had a simple bilinear fermionic action.
The ground states also factorize into direct products of one
factor for each chiral component. However, when the gauge field
background carries nontrivial topology, the ground states, which
for perturbative fields are singlets under the global chiral
group, carry nontrivial charges. By this mechanism global anomalous
conservation laws behave as expected (i.e. the respective charges
are not conserved, in spite of the formal decoupling in the action).
This property is needed in QCD, as is well known. 

In any regularization,
as long as the fermionic action is bilinear, the chiral components
either decouple or not. If we have a lattice model with a finite
number of fields per unit Euclidean volume one cannot have exact
chiral symmetries without this decoupling and one cannot get the
violations of anomalous conservation laws if one has exact
decoupling. The overlap's way
out is to be equivalent to a system containing an {\sl infinite} number
of fermions (heavy flavors). 

Recently, a particular truncation of the number of heavy flavors
has been applied to numerical QCD [\blumsoni] in four dimensions,
to the two dimensional two flavor Schwinger 
model [\jaster,\vranas] and compared
to the overlap in [\vranas]. The fermions in this
truncation are typically referred to as domain wall fermions.
This terminology is a residue of a very influential paper by Kaplan
[\kaplan],
who started this whole subset of activity in lattice field theory.
The fermions used in [\blumsoni], [\jaster] and [\vranas] are also sometimes
referred to as Shamir fermions [\shamir,\fursham]. They are slightly
different from the original ``Kaplan fermions'' in that they correspond
to a particular limiting case where a certain unimportant
free mass parameter is taken to infinity [\nninfmass]. This makes one of the
two ground states needed for the overlap
trivial and independent of the gauge field background, leaving only
the other ground state as the carrier of all the information typically residing
in the closed fermion loop vacuum diagrams. 

The main approximating feature of these systems is that one uses only a finite
number of fermions and nevertheless expects to get an essentially
chirally symmetric theory. As emphasized in [\vranas] it is important
then to compare carefully the truncated version to the overlap.
This was done numerically in [\vranas] for a toy model. 
On the
other hand, recent numerical work in the truncated model [\blumsoni]
for QCD produced promising results. The simplicity of the
main formula in [\recent] indicates that the comparison first undertaken
in [\vranas] at the numerical level can be attempted also at the
analytical level. In view of the work in [\blumsoni] such an analytical
comparison is needed, given the difficulty to simulate the overlap
directly in four dimensions. This leads us to the purpose of this paper,
namely, to 
improve our understanding of 
the nature of the approximation introduced by the truncation and of
its limitations. 

\figure{1}{\captionone}{3.5}
\figure{2}{\captiontwo}{3.5}
\figure{3}{\captionthree}{3.5}

To get a visual image for what the overlap does and how the truncations
approximate that consider the three schematic phase diagrams in figures
1,2,3. The structure of the phase diagrams is basically guessed and the
guess is quite incomplete by itself. Probably, 
numerical simulations of QCD would benefit from an investigation
of the phase diagram as a whole, at least in the regime of $\beta$-gauge
couplings that are practically relevant. 

The continuum limit is obtained as $\beta$ is taken to infinity. When
$\beta$ is small the link matrices fluctuate strongly, lattice
effects are important and the whole concept of chirality looses 
its meaning. m is a parameter that controls the dominating
correlation length among the fermions when the gauge forces
are turned off. The light fermion has a mass that decreases
with m in that case. When gauge interactions are turned on 
we choose to preserve CP invariance and keep m real. In the
continuum the physics for positive and negative m could 
differ substantially.

In the overlap (figure 1) there is a region where 
one has exact chiral symmetry at finite lattice spacing. To 
simulate numerically QCD one only needs chiral symmetries (ignoring
for the moment the nonzero light quark masses) in the continuum
limit, that is at $\beta=\infty$. The advantage of having chiral symmetry
at finite $\beta$ is that the approach to continuum is faster
as the leading lattice  scaling violations come, in the case of Wilson
fermions, from operators of dimension five, which also break
chiral symmetries. Eliminating chiral symmetry breaking also eliminates
these operators. In contemporary parlance the overlap
provides automatic ``nonperturbative ${\cal O} (a)$ improvement''. 
In practice this has been seen to work
in two dimensions (see Fig 1 of [\kiknn]; footnote 3
on page 110 there notes the importance for QCD). 

When $\beta$ is decreased, the links behave more or less
like $rU$ 
where $r$ is a positive real number less than one, decreasing
towards zero, and $U$ is unitary. This induces a reduction in the range
of masslessness, until, it is conjectured, the range shrinks to zero
at the point A. For coupling constants $\beta$
below $\beta_A$ there are no massless
quarks any more. The mesons are likely also all heavy, except
the possibility of a Wilson critical line. One manifestation
of the regime above ${1\over \beta_A}$ is the 
absence of instantons as detected
by the overlap fermions [\npblong]. $\beta_A$ is not known at present,
but, for $SU(2)$ in the quenched approximation, it is apparently
smaller than values of $\beta$ of numerical interest [\narvra].
Therefore, usage of the overlap appears viable in four dimensions
(QCD) even with presently available computing power. 

To the left of the line AB one expects different sets of degenerate
doublers to become massless. At $k=\infty$ there likely are more
transition lines there. Our discussion below 
ignores this region of parameter
space as it is quite unclear whether it would be of practical use
in simulations.

With the truncation, exact chiral symmetries are lost and dimension five
operators come back in. For large enough numbers of heavy flavors the
quarks get small masses, and the coefficients of the dimension five
operators are likely small numbers. There no longer are sharp
demarcation lines connecting A to B and A to C. These lines are replaced by
crossovers. However, a new critical line appears connecting
A to the origin. This is just the ordinary Wilson critical line.
As the number of heavy flavors is further reduced the standard
Wilson situation is approached. 

No claim is made that the above sketches are completely correct.
It is hoped that they do
capture some essential features that we should keep in mind when
reading the rest of the paper. In the next section the 
effective action will be discussed and some new results will be
presented. The technical device consists of some determinant formulae
derived in the appendix. The following section presents various
explanations of the mechanism that keeps the regulated theory
close to a chiral limit. The final section contains some conclusions
and suggestions for further research.

\vskip .5in

\centerline{\bf 2. Effective Action}
Using some manipulations on determinants we derive an expression
for the effective action induced by integrating over all fermions
in the truncated model. Our objectives are:
\item{$\bullet$} Make no direct use of operator formulae. The
introduction of the auxiliary Hilbert space is necessary in the
infinite flavor case [\npold]. It has been reused in [\fursham]
for the truncated case where it is not necessary, and obscures
the simple fact that all we are doing is studying lattice
QCD with several flavors mixed in a certain way. Once we understand
this it should come as no surprise that the mechanism for
suppressing one of the quark masses is well known in ordinary
continuum field theory and conceptually requires no extra
dimensions. 
\item{$\bullet$} We wish to make direct contact with the exactly
massless case [\recent] and see how close we would get to
it for typical gauge backgrounds.
\item{$\bullet$} The formulae provide the starting point for finding
expressions for the fermion correlation functions.
\item{$\bullet$} The overlap formulation is related to the
path integral one by a subtraction removing the effects of
most of the heavy quarks [\plbfirst,\npold]. 
The expression for the effective
action in [\recent] includes this subtraction. It is crucial
to carry the subtraction out correctly if one wants
to reproduce instanton effects. The subtraction of [\plbfirst]
was mentioned in [\shamir] but was viewed as unessential.
The adaptation of the prescription of [\plbfirst] to
the truncated case in [\fursham] is not quite right,
as first pointed out in [\vranas]. We show explicitly
that the subtraction in [\vranas] 
is the most natural one. 
\item{$\bullet$} The expressions we arrive at will allow
the introduction of two kinds of mass terms, and will 
establish that a proposal to look at the spectrum of the
operator $1+V$ in [\recent] could indeed show how spontaneous
chiral symmetry breakdown occurs, separating cleanly this
effect from those of nonzero global gauge field topology.

In [\plbfirst] an infinite mass matrix was introduced which
was shown to be equivalent to Kaplan's [\kaplan] fifth
dimension formulation. There were two parameters ($a_\pm$
in the notation of [\plbfirst]) of opposite signs that were
needed. While one was bounded, the other was not, and could
be taken to infinity. This results in a simplification
[\nninfmass] and is equivalent to so called
``open boundaries'' from the fifth dimensional viewpoint.
These open boundaries have appeared in the literature before,
in the present context in [\boya], but, according to [\creutzbnd]
also much earlier [\shotkey]. As an application to vector
like theories open boundaries were first
employed in [\shamir] and [\fursham].

The fermionic actions we shall look at are of the form:
$$
S=-\sum_{s=1}^{2k} \bar\Phi_s (D\Phi )_s .\eqno{(2.1)}$$
The fermionic fields, $\Phi_s$, have the following left-right structure:
$$
\pmatrix{ \Phi_1 \cr \Phi_2 \cr \vdots\cr\vdots\cr\Phi_{2k-1}\cr
\Phi_{2k}\cr} =
\pmatrix{ \chi_1^R \cr \chi_1^L \cr \vdots\cr\vdots\cr\chi_{k-1}^R\cr
\chi_{k}^L\cr} .\eqno{(2.2)}$$
$\chi_j^{R,L}$ are left or right Weyl fermions in the notation of
[\npold]. Similarly one defines $\bar\Phi_s$. Our convention is
that vector gauge interaction appear diagonal
in $D$. We suppress all space-time, spinorial and gauge indices, displaying
explicitly only the left-right character and flavor.

The lattice is taken to have $L^d$ sites. Our basic building blocks in the
matrix will have size $q\times q$ where, in $d$ dimensions (depending
on context, we shall often implicitly 
assume $d=4$) $q=2^{{d\over 2}-1} n_c L^d$.
$n_c$ is the dimension of the gauge group representation ($n_c =3 $
for QCD). Following [\npold] we write:
$$
D=\pmatrix{ C^\dagger & B & 0 & 0 & 0 & 0 &\ldots &\ldots & 0 & 0\cr
            B 	     & -C & -1& 0 & 0 & 0 &\ldots &\ldots & 0 & 0\cr
	    0   &-1& C^\dagger& B & 0 & 0 &\ldots &\ldots & 0 & 0\cr
            0 & 0  &     B    &-C &-1 & 0 &\ldots &\ldots & 0 & 0\cr
	    0 & 0  & 0 &-1& C^\dagger & B &\ldots &\ldots & 0 & 0\cr
            0 & 0  & 0 & 0&  B        &-C &\vdots &\ldots & 0 & 0\cr
            \vdots&\vdots&\vdots&\vdots&\vdots&\ddots&\ddots&\ddots
&\vdots&\vdots\cr
            \vdots&\vdots&\vdots&\vdots&\vdots&\vdots&\ddots&\ddots
&\ddots&\vdots\cr
	    \ldots&\ldots&\ldots&\ldots&\ldots&\ldots&\ldots&\ddots
&\ddots&\ldots\cr
	    0 & 0 &  0 & 0& 0         & 0 &\ldots &\ldots & B &-C\cr}.
\eqno{(2.3)}$$
The matrix $D$ is of size $2k\times 2k$ where the entries 
are $q\times q$ blocks. 

The matrices $B$ and $C$ are dependent on the gauge background defined
by the collection of link matrices $U_\mu (x)$. These matrices
are of dimension $n_c\times n_c$. $\mu$ labels the positive $d$ directions
on a hypercubic lattice and  $U_\mu (x)$ is the unitary matrix associated
with a link that points from the site $x$ in the $\hat \mu$-direction. 
$$
\eqalign{
( C )_{x \alpha i, y \beta j}
& ={1\over 2} \sum_{\mu=1}^{d} \sigma_\mu^{\alpha\beta}
[\delta_{y,x+\hat\mu} (U_\mu (x) )_{ij} -
\delta_{x,y+\hat\mu} (U_\mu^\dagger (y))_{ij}] \equiv 
\sum_{\mu=1}^{d} \sigma_\mu^{\alpha\beta} ( W_\mu )_{xi,yj}~~ \cr
( B_0 )_{x \alpha i, y \beta j} & = {1\over 2} \delta_{\alpha\beta} 
\sum_{\mu =1}^{d} [2\delta_{xy}
\delta_{ij} - \delta_{y,x+\hat\mu} (U_\mu (x) )_{ij} -
\delta_{x,y+\hat\mu} (U_\mu^\dagger (y))_{ij} ]\cr
( B )_{x \alpha i, y \beta j} & =( B_0 )_{x \alpha i, y \beta j} 
+ 
M_0 \delta_{x\alpha i,y\beta j}\cr}\eqno{(2.4)}$$
The indices $\alpha,\beta$ label spinor indices in the range $1$ 
to $2^{{d\over 2}-1}$. The indices $i,j$ label color in 
the range $1$ to $n_c$. The Euclidean $2^{d\over 2} \times 2^{d\over 2}$
Dirac matrices $\gamma_\mu$
are taken in the Weyl basis where their form is
$$
\gamma_\mu = \pmatrix {0&\sigma_\mu \cr \sigma_\mu^\dagger & 0\cr}.
\eqno{(2.5)}$$

Of particular importance is the
parameter $M_0$. As long as $M_0 > 0$ the matrix $B$ is positive
definite due to the unitarity of the link variables.
To make almost massless quarks on the lattice one also
wants [\kaplan, \plbfirst] $|M_0|<1$. 
(Note that the notational conventions adopted here are
slightly different from [\kaplan] and 
[\plbfirst]: The parameter $M_0$ often appears
as $1-m_0$ and the parameter m in figures 1,2,3 although
meant there more generically, is just $M_0$ here.) 
Although several
of the manipulations require that $B$ be nonsingular,
and therefore one would restrict $M_0$ to 
the interval $(0,1)$ (this is probably an overkill)
the final expressions are meaningful
for the entire range $(-1,1)$. 

It is convenient to introduce the $2q \times 2q$ matrix $\Gamma_{d+1}$
representing the regular $\gamma_{d+1}$ matrix on spinorial indices
and unit action on all other indices. In terms of $q\times q$ 
blocks we have
$$
\Gamma_{d+1} =\pmatrix{1&0\cr0&-1}.\eqno{(2.6)}$$

Setting $X=Y=0$ in eq. (A.15) we get:
$$
\det D = (-)^{qk} (\det B )^k \det \left [ {{1-\Gamma_{d+1}}\over 2}
-T^{-k}  {{1+\Gamma_{d+1}}\over 2}\right ].
\eqno{(2.7)}$$
Writing
$$
T\equiv e^{-H} = \pmatrix {{1\over B} & {1\over B} C \cr
         C^\dagger {1\over B} & C^\dagger {1\over B} C +B \cr},
\eqno{(2.8)}$$
we obtain 
$$
\det D = (-)^{qk} (\det B )^k \det \left [1 +e^{kH}\right ]
\det \left [ {{1+\Gamma_{d+1} \tanh ({k\over 2} H )}\over 2} 
\right ].
\eqno{(2.9)}$$

Comparing to the overlap formula in [\recent], 
we conclude that we want the subtraction to remove all factors but
the last in the above equation. As we shall see below, the 
large $k$ limit is then precisely given by the overlap. The factors
we wish to cancel out exactly correspond to the determinant induced
by integrating over the fermions of 
a system identical to the ones we dealt
with up to now, only that the boundary conditions at $s=1$ 
and at $s=2k$ have to be
chosen as {\sl anti-periodic}. This should come as no surprise,
since the factor $\det \left [1 +e^{kH}\right ]$ above clearly
corresponds to the {\sl trace} of the transfer matrix $T$ and a trace
is implemented by anti-periodic boundary conditions when the integration
variables are Grassmann. 

The subtraction is handled by
adding $2k$ pseudo-fermions, i.e. fields that have identical 
index structure to fermions, only their statistics is assumed to be
of Bose type. The pseudo-fermions are coupled by a matrix $D^{pf}$.
If the subtraction is implemented in a Monte Carlo
simulation it will be important that the pseudo-fermion
determinant not change sign as a function of the gauge background.
$$
D^{pf} =
	 \pmatrix{ C^\dagger & B & 0 & 0 & 0 & 0 
&\ldots &\ldots & 0 & 1\cr
            B 	     & -C & -1& 0 & 0 & 0 &\ldots &\ldots & 0 & 0\cr
	    0   &-1& C^\dagger& B & 0 & 0 &\ldots &\ldots & 0 & 0\cr
            0 & 0  &     B    &-C &-1 & 0 &\ldots &\ldots & 0 & 0\cr
	    0 & 0  & 0 &-1& C^\dagger & B &\ldots &\ldots & 0 & 0\cr
            0 & 0  & 0 & 0&  B        &-C &\vdots &\ldots & 0 & 0\cr
            \vdots&\vdots&\vdots&\vdots&\vdots&\ddots&\ddots&\ddots
&\vdots&\vdots\cr
            \vdots&\vdots&\vdots&\vdots&\vdots&\vdots&\ddots&\ddots
&\ddots&\vdots\cr
	    \ldots&\ldots&\ldots&\ldots&\ldots&\ldots&\ldots&\ddots
&\ddots&\ldots\cr
	    1& 0 &  0 & 0& 0         & 0 &\ldots &\ldots & B &-C\cr}
\eqno{(2.10)}$$
Setting $X=Y=1$ in equation (A.15) we get:
$$
\det D^{pf} = (-)^{qk} (\det B )^k \det \left [1 +e^{kH}\right ].
\eqno{(2.11)}$$
Sign changes would be avoided if $B$ is a positive matrix. 

We now obtain the effective action for the truncated model:
$$
{{\det D}\over {\det D^{pf}}} = 
\det \left [ {{1+\Gamma_{d+1} \tanh ({k\over 2} H )}\over 2} 
\right ]\eqno{(2.12)}$$
To connect to the overlap we let $k\to\infty$ and obtain:
$$
 {{\det D}\over {\det D^{pf}}}\to  
\det \left [ {{1+\Gamma_{d+1} \epsilon (H )}\over 2} 
\right ].
\eqno{(2.13)}$$
This formula is identical in structure to the main result of [\recent].
There is a difference though: Here $H$ is more complicated and not
strictly local. The difference reflects the usage of discrete flavor
here as opposed to continuous flavor in [\recent]. With continuous
flavor $H$ simplifies significantly. In a simulation however, discrete
flavor is more appropriate. On the basis of the above we can 
also write down the effective action for the truncated system 
in the continuous flavor case:
$$
\left ( {{\det D}\over {\det D^{pf}}}\right )^\prime  = 
\det \left [ {{1+\Gamma_{d+1} \tanh (\Delta H^\prime )}\over 2} \right ].
\eqno{(2.14)}$$
The parameter $\Delta$
represents the finite range of continuous flavor.
In the overlap, $\Delta$ is taken to infinity. 
The matrix $H^\prime$ above is simply related to a Wilson-Dirac 
lattice operator, $X$:
$$
\Gamma_{d+1} H^\prime \equiv X = \pmatrix {B^\prime & C\cr
-C^\dagger & B^\prime }.\eqno{(2.15)}$$
The block $B^\prime$ is the same as $B$ introduced before,
only the parameter $M_0$ is taken to the range $(-2,0)$. 
Therefore, for continuous flavor, we deal only with {\sl negative}
masses, while, for discrete flavor, from the $d$-{\sl dimensional point
of view}, we can restrict ourselves to positive $M_0$ only.
From the $d+1$-dimensional point of view one sometimes adopts
if one thinks in terms of domain walls one would say that one
always has a negative mass term. But, as far as questions
of positivity go, the $d+1$ terminology is misleading.

Turning back to discrete flavor we arrive at a criterion for when the
truncated expression is a good approximation to the exactly massless
system:
$$
k >> \max_t \Big | {1\over{log|t|}} \Big |
\eqno{(2.16)}$$
The maximum is taken over all 
eigenvalues of $T$, $t$, where $T$ is given by equation (2.8). 
As long as the gauge configurations are smooth in the gauge
invariant sense, one expects a gap in the spectrum of $T$
around $1$ and the criterion is not very restrictive. For large
enough gauge coupling $\beta$ one expects such configurations
to dominate. Thus, if we had a method of generating gauge configurations,
each one from scratch and correctly distributed, the criterion
might end up to be satisfied in practice for all configurations
with a reasonably small $k$ (in the last
section some suggestion for lowering the needed $k$ even further are made).
But, the real life simulation methods are based on a walk in
the space of gauge configurations. Thus, gauge configurations
evolve at some rate in the space. This evolution has to
produce, with the correct probability, configurations that
approximate continuous backgrounds carrying nontrivial
topological charge. During the evolution between two configurations
that carry different topological charge the gauge fields
must pass through points where they are very different
from a smooth background in the sense that they have,
at some location, a structure that would be interpreted in the continuum
as a singularity. In the vicinity of those
configurations the matrix $T$ must have one eigenvalue at least
that is very close to unity. This is so because the number
of eigenvalues of $T$ that is smaller than unity changes
when one goes from a configuration that carries one topological
charge to a configuration that carries another
(the geometric mean of the eigenvalues of $T$ is fixed
by $\det T =1$). This is
a problem noted in [\vranas] on the basis of experimentation
with the two flavor Schwinger model. While it is essentially
an algorithmic problem, it is an old one, so an immediate
clean resolution is not very likely. 

For any finite $k$ the partition function never vanishes and,
actually, stays positive. In the infinite $k$ limit robust zeros
appear in instanton backgrounds. For this, accurate subtraction is 
essential. The interpretation of the finite $k$ system
is that it contains $k-1$ heavy quarks and one light quark
(we ignore the heavy doublers here). The light
quark is almost massless when $k$ is large. The mass of the light
quark is {\sl positive} and vanishes exponentially as $k$ increases.

Although the light fermion has a small mass for any finite $k$, one
may want to add yet another mass term, $\mu$. In particular, one may
wish to study what happens when the mass is allowed to go negative
(more precisely, what happens when, for small $\mu$, 
the combination $\mu e^{i\theta}$ goes negative  - here $\theta$ 
is the famous theta-parameter). 

The simplest way to add another mass parameter is to add a direct
coupling between the left and right components of the would be
massless quark. The most natural interpolating fields for the
massless quark were defined in [\plbfirst]: They are the fields
at the ``defect'' in Kaplan's picture. The appropriate mass term
was introduced in [\npblong] and shown to have (assuming its
sign is positive) the needed properties to ensure 
Nussinov-Weingarten-Witten mass inequalities [\massineq]
at the regularized level. 

To identify the left and right components of the would be massless
quark in this case is easy: Set $B=0$ in the expression for $D$.
It is evident then that one has at $s=1$ a massless right handed
fermion and at $s=2k$ a massless left handed fermion. In addition
there are $k-1$ massive Dirac fermions. Actually, every one of the
Dirac fermions comes in $2^d$ copies because of lattice doubling. $B$ 
takes care of the extra massless $2^d -1$ copies and makes them 
heavy. $B$ also couples the two remaining 
light Weyl fermions, but the coupling
is indirect and only a small mass is generated. This mass vanishes
as $k\to\infty$. To maintain a mass also in this limit, similarly
to [\npblong] we introduce $\mu$ via the matrices $X$ and $Y$ of the
appendix. 

The new operator acting on the fermions is then $D(\mu )$ given by:
$$
D(\mu ) =\pmatrix{ C^\dagger & B & 0 & 0 & 0 & 0 &\ldots &\ldots & 0 & \mu\cr
            B 	     & -C & -1& 0 & 0 & 0 &\ldots &\ldots & 0 & 0\cr
	    0   &-1& C^\dagger& B & 0 & 0 &\ldots &\ldots & 0 & 0\cr
            0 & 0  &     B    &-C &-1 & 0 &\ldots &\ldots & 0 & 0\cr
	    0 & 0  & 0 &-1& C^\dagger & B &\ldots &\ldots & 0 & 0\cr
            0 & 0  & 0 & 0&  B        &-C &\vdots &\ldots & 0 & 0\cr
            \vdots&\vdots&\vdots&\vdots&\vdots&\ddots&\ddots&\ddots
&\vdots&\vdots\cr
            \vdots&\vdots&\vdots&\vdots&\vdots&\vdots&\ddots&\ddots
&\ddots&\vdots\cr
	    \ldots&\ldots&\ldots&\ldots&\ldots&\ldots&\ldots&\ddots
&\ddots&\ldots\cr
	    \mu & 0 &  0 & 0& 0         & 0 &\ldots &\ldots & B &-C\cr}.
\eqno{(2.17)}$$
Clearly, the pseudo-fermion operator, $D^{pf}$ is just $D(\mu =1 )$.
Therefore, the subtraction simply removes $k$ massive Dirac fermions
(not counting doublers). Note that one removes $k$ heavy particles
although the original system had only $k-1$ heavy particles. In this
sense, the remaining system can be thought of as ``doubly'' regularized:
in addition to the lattice one also has a 
Pauli-Villars regulator.
Here, we draw a distinction between the ``latticy''
concept of a pseudo-fermion and
the continuum concept of a Pauli-Villars one. In [\shamir, \fursham, \vranas]
any fermion with wrong statistics is referred to as a Pauli-Villars
fermion. These papers also include a $\mu$-mass term. 

Manipulations similar to ones already employed yield:
$$
\det D(\mu ) = (-)^{qk} (\det B)^k \det \left [ 1 + e^{kH} \right ]
(1+\mu )^q \det {1\over 2} \left [ 1+ \Gamma_{d+1} {{1-\mu}\over{1+\mu}}
\tanh \left ( {k\over 2} H \right ) \right ].
\eqno{(2.18)}$$
Clearly, we take $|\mu|< 1$ with $0<\mu<1$ representing positive masses
and $-1 < \mu < 0$ representing negative masses, where, by definition,
we take $\theta=0$. Note that, just like in [\npblong], the subtraction
of pseudo-fermions 
does not depend on $\mu$. This is important, since the determinant
associated with the subtraction must have a definite sign, while
the determinant of the original fermions must change sign for negative
masses when the gauge fields change topology by one unit. 
Let us remark that using the general 
matrices $X$ and $Y$ of the appendix one
can similarly
derive left-right correlation functions for the 
light quark by differentiation.
This bypasses the need for operators. 

To see the relation between topology and mass sign come out we ought to
dispense with the other source of finite mass, namely $k$. We go
to the overlap then, setting $k=\infty$, and obtain:
$$
d(\mu )\equiv \left [ {{D(\mu )}\over {D(\mu =1)}}\right ]_{k=\infty} =
\det {1\over 2} \left [ 1 +\mu +(1-\mu ) 
\Gamma_{d+1} \epsilon (H) \right ]. \eqno{(2.19)}$$
Assuming $\det H \ne 0$ we obtain:
$$
d(\mu ) = \det \left [ \Gamma_{d+1} \epsilon  (H) \right ] d (-\mu ).
\eqno{(2.20)}$$
In [\npold, \recent] we defined the 
topological charge $n_{\rm top}$ as half
the difference between the number of positive and negative eigenvalues
of $H$. This implies
$$
d(\mu ) = (-)^{n_{\rm top}} d (-\mu ) ,\eqno{(2.21)}$$
confirming the interpretation of the sign of $\mu$ as the sign of the
physical mass. Similarly, one could treat complex masses, or (in the 
case of several light flavors) mass matrices. The simplicity of the
formula raises the hope of possible lattice investigations of $QCD$
at $\theta = \pi$. 

We immediately learn now how to add a mass term to the continuum flavor
case where $\mu$ is the {\it single} source for a mass: all we need
to do is use in $d(\mu)$ 
the simpler matrix in [\recent] which plays the role of
$H$ here (we denoted this matrix by
$H^\prime$ in (2.14-15)).
Also, with $V=\Gamma_{d+1} \epsilon (H^\prime)$, we see, by taking derivatives
with respect to $\mu$ at $\mu=0$, that spontaneous chiral symmetry
breakdown should indeed be found in the spectral properties of $1+V$,
establishing the validity of a conjecture in [\recent].

Indeed, for $N_f$ degenerate flavors $f$ we have:
$$
{1\over {N_f}}
\sum_{f=1}^{N_f} < (\bar\psi_f \psi_f )_{\rm physical} > =
{{\cal Z}\over{ L^d}} <{\det}^{N_f} 
\left [ {{1+V}\over 2}\right ] {\rm Tr} {{1-V}\over{1+V}}
>_{\cal A}.\eqno{(2.22)}$$
Here $<..>_{\cal A}$ means an average with respect to the pure gauge action
and ${\cal Z}$ is a renormalization constant. We see how single
instantons would give a nonzero contribution for one flavor,
but no contribution for more flavors, just as expected from
the more formal continuum expressions.

Note that the factor that appears traced is ${{1-V}\over{1+V}}$ rather
than just ${2\over{1+V}}$, but since the interesting regime is at
eigenvalues of $V$ close to $-1$ the difference can be absorbed
in ${\cal Z}$.  Let us look at the
difference between ${{1-V}\over{1+V}}$ and
${2\over{1+V}}$ in the free case\footnote{*}{This is an
observation made by Ting-Wai Chiu 
in a private communication to the author.}:
Write, for the free case,
$$
V= {{i\gamma_\mu Q_\mu + M}\over {\sqrt{Q^2 + M^2 }}},\eqno{(2.23)}$$
where $ Q_\mu$ and $M$ can be function of the momenta $p_\nu$.
One gets then,
$$
(1-V){1\over{1+V}} = (1-V) {{1-V^\dagger}\over {V-V^\dagger}}=
{{2-V-V^\dagger}\over{V-V^\dagger}}=
{{\sqrt{Q^2 +M^2 } -M}\over {i\gamma_\mu Q_\mu }}.
\eqno{(2.24)}$$
This expression is close to the continuum in that it anticommutes
with $\gamma_5$. $M$ is chosen to be nonzero and positive at the
location of all four-momenta which make $Q_\mu =0$ except the
zero four-momentum point where $M$ is negative. The elimination
of the doublers is of the same type as first proposed by Rebbi [\rebbi].
Since ${2\over {1+V}} = {{1-V}\over{1+V}} +1 $ the former
expression has a remnant of chiral symmetry of the form discussed
some time ago by Ginsparg and Wilson [\ginspwil]. 

To be sure,
neither expression appears to be a complete and unique
replacement for the
massless continuum fermion
propagator (in the presence of gauge fields).
Actually, it is not certain that such an object 
really exists or is at all necessary. 
Of course, there does exist a full fermion
propagator, including all heavy fields in addition to the
physical field. This propagator can be 
obtained directly from the $D$ matrix. However,
if we wish the propagator for, say the chiral components
at $s=1$ and $s=2k$ all other fermions need to be integrated
over, just like in [\plbfirst]. The 
relevant expression 
could be obtained by using appendix A to couple external sources to the
desired bilinears, and differentiating subsequently the 
exact formulae for the determinants with respect to the sources.
From the expressions we derived until now 
we learn that for the purpose of computing the
effective action we can use ${2\over {1+V}}$ as a propagator
of the single physically interesting fermion, 
while, for the $<(\bar\psi \psi )_{\rm phys}>$ condensate, the
role of the physical propagator factor is played by ${2\over {1+V}}-1$,
although, the determinant factor is still $\det {{1+V}\over 2}$.

Supersymmetric theories with no chiral matter contain fermions
in real representations of the gauge group. The simplest case is
theories with no matter at all. As emphasized by Curci and Veneziano
[\curciv] supersymmetry should be restored in the continuum limit
if enough ordinary symmetries are preserved on the lattice. Thus,
as noted in [\npblong], the overlap could be used for 
supersymmetric theories. Here we wish to present simplified
formulae
of the type in [\recent] for 
the effective actions
for the supersymmetric case, both in the truncated models
and in the overlap
limit. For other work with Majorana fermions see [\aokizen].

In the supersymmetric case of pure gauge we have fermions in a real
representation so the $U_\mu$ matrices are real. Let us restrict
our attention to $d=4$ for definiteness. What we need then is
a square root of the determinant that is {\bf analytic} in the
link variables. The analyticity in the link variables
assures that no unwanted new terms would come into Ward identities
involving the link matrices: If, for example, we just took the
square root of the absolute value of the determinant, in the
vicinity of configuration for which the determinant vanishes
we could generate delta-functions of $U_\mu$ variables
if we take derivatives with respect to $U_\mu$ a sufficient number of
times. Such terms may spoil a smooth approach to the continuum
limit, by refusing to disappear. Following [\huetnn] we are looking
for a decoupling of the fermionic functional integral in the case
of Dirac fermions in a real representation of the gauge group. 
Such a decoupling is achieved 
if one finds a basis in which the matrix still has 
entries analytic in the $U_\mu$ but where it is antisymmetric.
The analytic square root is then the pfaffian. 

We first make a simple basis change, reversing the order of the
rows of $q\times q$ blocks in the matrix $D(\mu )$.

$$
D_r (\mu ) =
\pmatrix{ \mu & 0 &  0 & 0& 0         & 0 &\ldots &\ldots & B &-C\cr
 \ldots&\ldots&\ldots&\ldots&\ldots&\ldots&\ldots&\ddots
&\ddots&\ldots\cr
\vdots&\vdots&\vdots&\vdots&\vdots&\vdots&\ddots&\ddots
&\ddots&\vdots\cr
 \vdots&\vdots&\vdots&\vdots&\vdots&\ddots&\ddots&\ddots
&\vdots&\vdots\cr
            0 & 0  & 0 & 0&  B        &-C &\vdots &\ldots & 0 & 0\cr
	    0 & 0  & 0 &-1& C^\dagger & B &\ldots &\ldots & 0 & 0\cr
            0 & 0  &     B    &-C &-1 & 0 &\ldots &\ldots & 0 & 0\cr
	    0   &-1& C^\dagger& B & 0 & 0 &\ldots &\ldots & 0 & 0\cr
            B 	     & -C & -1& 0 & 0 & 0 &\ldots &\ldots & 0 & 0\cr
C^\dagger & B & 0 & 0 & 0 & 0 &\ldots &\ldots & 0 & \mu\cr  }.\eqno{(2.25)}
$$

Clearly,
$$
\det D_r (\mu ) = (-)^{kq} \det D (\mu ).\eqno{(2.26)}$$

Because of the reality of the representation, $U_\mu (x) = U_\mu^* (x)$
for all $\mu$ and $x$. The $B$ block is therefore real, while the $C$
block obeys
$$
C=\sum_{\mu =1}^d \sigma_\mu W_\mu ,\eqno{(2.27)}$$
where $W_\mu =-W_\mu^T=W_\mu^*$ with $W^T$ meaning the transpose of $W$.
In four dimensions we choose $\sigma_4 =i$ and $\sigma_{1,2,3}$ as 
the standard Pauli matrices. 
Therefore $\sigma_2 \sigma_\mu \sigma_2 = -\sigma_\mu^*$. We learn that
$$
\eqalign{ \sigma_2 C^T \sigma_2 =& - C^\dagger  \cr
 	  \sigma_2 B \sigma_2 =& B^T .\cr}\eqno{(2.28)}$$
Define a block diagonal $2k \times 2k$ matrix $\Sigma_2$ with $q\times q$
blocks given by $\sigma_2$ on spinorial indices and unity on all
other indices. The matrix we are interested in in the supersymmetric case
is $D_{\rm SUSY} (\mu ) = \Sigma_2 D_r (\mu)$,
$$
D_{\rm SUSY} (\mu ) =
\pmatrix{\mu \sigma_2  & 0 &  0 & 0& 0         & 0 &\ldots &\ldots 
& B^\sigma &-C^\sigma\cr
 \ldots&\ldots&\ldots&\ldots&\ldots&\ldots&\ldots&\ddots
&\ddots&\ldots\cr
\vdots&\vdots&\vdots&\vdots&\vdots&\vdots&\ddots&\ddots
&\ddots&\vdots\cr
 \vdots&\vdots&\vdots&\vdots&\vdots&\ddots&\ddots&\ddots
&\vdots&\vdots\cr
            0 & 0  & 0 & 0&  B^\sigma        &-C^\sigma 
&\vdots &\ldots & 0 & 0\cr
	    0 & 0  & 0 &-\sigma_2 & C^{\sigma T} 
& B^\sigma &\ldots &\ldots & 0 & 0\cr
            0 & 0  &     B^\sigma    &-C^\sigma 
&-\sigma_2 & 0 &\ldots &\ldots & 0 & 0\cr
	    0   &-\sigma_2 & C^{\sigma T}
& B^\sigma & 0 & 0 &\ldots &\ldots & 0 & 0\cr
            B^\sigma 	     & -C^\sigma & -\sigma_2& 0 
& 0 & 0 &\ldots &\ldots & 0 & 0\cr
C^{\sigma T} & B^\sigma & 0 & 0 & 0 & 0 &\ldots &
\ldots & 0 & \mu\sigma_2\cr  }.\eqno{(2.29)}
$$
Here,
$$
C^\sigma = \sigma_2 C , ~~~ C^\dagger = \sigma_2 C^{\sigma T} ,
~~~ B^\sigma = \sigma_2 B = -B^{\sigma T}.\eqno{(2.30)}$$
Since $\det \sigma_2 = -1$,
$$
\det D (\mu) = (-)^{kq} \det D_{\rm SUSY}(\mu ).\eqno{(2.31)}$$

Since $ D_{\rm SUSY} (\mu)$ is antisymmetric, 
we obtain, for the supersymmetric
case, the analytic square roots $pf  \left [ D_{\rm SUSY} (\mu )\right ]$. 
This implies that the ratio of determinants in the
subtracted system also admits an analytic square root,
and the effective action in the ``truncated'' supersymmetric case
is:
$$
\pm {{pf \left [ D_{\rm SUSY} (\mu =0 )\right ] }\over
{pf \left [ D_{\rm SUSY} (\mu =1 )\right ] }} \equiv
\sqrt {\det \left [ {{1+\Gamma_{d+1} \tanh ({k\over 2} H )}\over 2} 
\right ]}.\eqno{(2.32)}$$
There is an overall sign
on the left hand side that we have not determined; this
is not important since this sign is independent of the gauge field
because on the right hand side we have an expression that will
not vanish for any set of links $U_\mu$. (The single way the sign on
the left hand side could depend on the gauge field would be for
the expression to vanish when one gauge field configuration is deformed
into another.) The reason the expression under the
square root on the right hand side cannot vanish is the $\tanh$
term, which ensures that the operator $\Gamma_{d+1} \tanh ({k\over 2} H)$
has norm less than one for any finite $k$. 
This is just another way to see
that the would be gluinos have a finite positive mass when $k$ is
finite. Thus, we shall not have exact supersymmetry for any
finite $k$ in the continuum limit. 

When $k$ is taken to infinity the norm
becomes one and zeros become possible (and actually expected to
occur in instanton backgrounds). But, since sign 
changes as a function of the gauge background were not 
possible for any finite $k$, sign changes are prohibited also
in the infinite $k$ limit. This is in agreement with the
expectation from continuum [\hsu]. Therefore, we expect
a supersymmetric theory in the continuum limit, when
the gauge coupling $\beta$ is taken to infinity and the
mass parameter $M_0$ is kept anywhere within the 
finite interval $(0,1)$. In practice, one will be
working at a finite $\beta$ and then the range for $M_0$
is different, as sketched in the introduction. Also,
unless some new simulation trick is discovered, we would be
working at a finite $k$, large enough that the effects of finite
gluino mass are negligible when compared to other sources of
statistical and systematical errors.

The left hand side of the above equation tells us that
the expression is not only not changing sign
but also is analytic (more precisely,
for finite lattices and finite $k$ it is a ratio of polynomials
in the link variables).

In the massless limit ($k=\infty$) and going over to continuous
heavy flavor space we obtain the following compact expression
for the fermionic determinant on the lattice (the equality holds 
up to an irrelevant overall sign):
$$
\sqrt {\det \left [ {{1+\Gamma_{d+1} \epsilon ( H^\prime )}\over 2} 
\right ]}=
pf \left [ {{\pmatrix
{\sigma_2&0\cr 0&\sigma_2} +V_{\rm SUSY}}\over 2}\right ].
\eqno{(2.33)}$$
The unitary and antisymmetric matrix $V_{\rm SUSY}$
is given by:
$$
V_{\rm SUSY}=Y{1\over {\sqrt { Y^\dagger Y}}}, 
~~~~Y=\pmatrix {\sigma_2 B^\prime &
\sigma_2 C \cr -\sigma_2 C^\dagger &\sigma_2 B^\prime \cr}=-Y^T 
\eqno{(2.34)}$$
The above is a direct generalization of results in [\odddim].

Lattice approaches to supersymmetry in different contexts
have been discussed also in [\huetnn, \montvay, \nishimura, \italiansusy].
Some of these papers employ ordinary Wilson fermions and there
the issue of positivity on the lattice (versus the one in the continuum
[\hsu]) becomes somewhat murky.\footnote{*}{I thank Istvan Montvay for
an e-mail discussion of this point.} 

In our equations the link matrices where taken in the adjoint representation.
In the case of $SU(n_c )$ these $(n_c^2 -1)\times(n_c^2 -1)$
matrices can be expressed in terms of the $n_c \times n_c$
link matrices in the fundamental representation as is well known
from elementary group theory. 
Such a representation is natural if one keeps the bosonic gauge part
in terms of link matrices in the fundamental. There is no reason
to do that for supersymmetric theories and it is more
natural to take the Wilson plaquette term also in the adjoint. 
If this is done, lattice gauge field configurations that approximate
nontrivial continuum $SU(n_c ) /Z(n_c )$ bundles on tori can be
dynamically generated. This may be important in the continuum limit 
[\eyalgomez], in particular if we view the approach towards
the continuum limit as taking place on a four-torus of fixed physical
size.

\vskip .5in
\centerline{\bf 3. Mass Matrix}
In this section we deal with the mass matrix in the 
absence of the parameter $\mu$. Our main objective is
to see that indeed one gets one light quark and $k-1$ heavy ones.
Actually, what we are really interested in, is to identify
the mechanism that is capable of preserving the large hierarchy
between the heavy fermions and the light one. Phrasing it this
way, we see that our problem is very similar to well known
problems faced in Particle Physics. It should come as no
surprise then, that we shall conclude that so called domain
wall fermions in the truncated context are nothing new. On the
other hand, by identifying the mechanism that preserves the
hierarchy {\sl at finite} $k$, we shall gain confidence in this
way of regularizing almost massless quarks on the lattice.
Moreover, usage of standard Particle Physics concepts is usually
advantageous in lattice work. 

Some workers [\blumsoni] seem to believe that the overlap is somehow
exclusively restricted to chiral gauge theories. Logically this is
almost impossible, and, indeed, 
a significant part of section 9
in [\npblong] was devoted to vector-like theories. It was shown
there that for $N_f$ massless flavors 
an $SU(N_f)_R \times SU(N_f)_L\times U(1)_{R+L} $ 
is preserved exactly
at finite lattice spacings, thus, not only exhibiting exact chiral
symmetries (evidence for exact masslessness of the quarks), but also
the explicit breaking of $U(1)_{R-L}$ induced by gauge topology.

The heart of the mechanism that preserves the zero mass of the quarks
on the lattice was identified in [\plbfirst]: In the infinite internal
space the mass matrix has a nontrivial index 
(associated with a quantum mechanical supersymmetry
in heavy flavor space) which must be stable
under small but finite  radiative corrections (there is a finite
ultraviolet cutoff). Essentially, there exist well localized 
states in the internal space which represent unremovable zero modes
and create the left and right components of the massless quark. 
When flavor space is truncated the index is lost and the massless
quark acquires a small mass. The question is what keeps it small
once gauge interactions are turned on, and the index is absent. 

Clearly, when the internal space is very large, but not infinite, one
would expect the mechanisms that do the job for the untruncated
situation to be still at work and prevent the mass of the light
quark from becoming too heavy. Otherwise, an unlikely
discontinuity in the mass spectrum would occur as internal space 
expanded to infinity. It was shown in [\plbfirst] that fermion
propagators in internal space were exponentially decaying in the
infinite directions and therefore perturbation theory could not generate
such a discontinuity. Going from infinite internal 
space to finite internal space changes the propagator somewhat [\shamir]
but nothing much can happen given the exponential decays mentioned above.
To obtain the effects of the truncation on the infinite flavor
space propagator of [\plbfirst] one could use the method of images.

So, it looks plausible that a large hierarchy can be preserved
in the truncated case. But, once we have discrete flavor and truncated
flavor space, it seems absurd to think about flavor as a space
approximating something infinite. In this context 
one often finds statements about the truncated model 
describing it as a domain wall fermion model with exact chiral symmetries
in the limit of an infinite distance between the walls. This creates
an illusion that the number of flavors is just a parameter like a mass, 
and we can take it at will wherever we want. Clearly, the number of
fields in any model is not a usual parameter and taking it to
infinity is not a simple procedure. So, then, what would be a better way
to think about the truncated model ? The answer is simple: it is a model
with a special mass matrix designed to preserve a large mass hierarchy.
It should be possible to understand how it works without referring
to infinite flavor limits. Once we gain this understanding we can 
trust the mechanism to work on the lattice. 

We shall look at the mass hierarchy from three new points of view:
The first employs the theory of orthogonal polynomials and is somewhat
mathematical. The second shows a connection to the well known see-saw
mechanism [\seesaw]. The third establishes a 
link to the Froggatt-Nielsen [\fn] mechanism.

The mass matrix $M$ is made up of all terms coupling left to right handed
fermions in $D$. By permuting indices, one brings $D$ into the more
convenient form below:
$$
D=\pmatrix{C^\dagger & M \cr M^\dagger & -C\cr}.\eqno{(3.1)}$$
Here $C,C^\dagger$ are of size $kq \times kq$ with trivial 
action in flavor space. $M$ is a truncation of the 
infinite mass matrix of [\plbfirst]. 
$$
M=\pmatrix{B&0&0&\ldots&0&0\cr
          -1&B&0&\ldots&0&0\cr
          0&-1&B&\ldots&0&0\cr
 \vdots&\vdots&\ddots&\ddots&\vdots&\vdots\cr
 \vdots&\vdots&\vdots&\ddots&\ddots&\vdots\cr
          0&0&0&\ldots&-1&B\cr}.\eqno{(3.2)}$$
Let us consider diagonal $B$ matrices (free fields in momentum basis)
and compute the bare masses. We introduce new notation,
replacing the blocks by numbers. This is 
permissible since the blocks are diagonal
now.
$$
M=\pmatrix{b & 0 & 0 & \ldots & 0 & 0\cr
           a & b & 0 & \ldots & 0 & 0\cr
           0 & a & b & \ldots & 0 & 0\cr
       \vdots&\vdots&\ddots&\ddots&\vdots&\vdots\cr
      \vdots&\vdots&\vdots&\ddots&\ddots&\vdots\cr
	0 & 0 & 0 & \ldots & a & b \cr}.\eqno{(3.3)}$$
The mass eigenvalues are extracted from $M^\dagger M$.
$$
M^\dagger M = (a^2 +b^2 ) {\bf 1} + 
2ab (J-{a\over b} N_k ).\eqno{(3.4)}$$
$J$ is a truncated Jacobi matrix. The full Jacobi matrix is tri-diagonal
and generates a sequence of orthogonal polynomials. In this case the orthogonal
polynomials are the Chebyshev polynomials of second kind. The matrix
$N_k$ contains information about the truncation. $N_k$ has all entries
vanishing except the $(k,k)$ entry which is equal to ${1\over 2}$. 
Similarly,
we define a matrix $N_1$ whose single nonzero entry is at $(1,1)$
where it is again equal to ${1\over 2}$. $MM^\dagger$ obeys a similar formula
as above, only $N_k$ is replaced by $N_1$. $J$ is defined by:
$$
2J=\pmatrix{0&1&0&\ldots\cr
            1&0&1&\ldots\cr
	    0&1&0&\ldots\cr
	    \vdots&\ddots&\ddots&\ddots\cr}.\eqno{(3.5)}$$
$M$ is not diagonalizable:
$$
[M^\dagger , M] =2 a^2 (N_1 - N_k ).\eqno{(3.6)}$$

When $k$ is taken to infinity and one requires square summability over
flavors, the term $N_k$ in the above equation gets replaced by zero. 
From the point of view of the light degrees of freedom this amounts
to keeping only one of the chiral components. By relabeling backwards
one sees that it is possible to keep only the other chiral component. 
The infinite $k$ limits we took on the determinants before keep 
{\sl both} components. 

Suppose only one chiral component is kept: For example, assume
that $M^\dagger$ has 
a zero mode $|0>$, while $M$ has no zero mode. There is a nontrivial 
index and we have $|<0|[M^\dagger , M]|0>| > 0$. 
We conclude that in order to keep only 
one chiral component, it is necessary 
(but not sufficient), that 
$tr[M^\dagger , M]=\pm 2a^2 \ne 0$ in the limit.  
Since for any finite
$k$,  $tr[M^\dagger , M] = 0$, these limits 
are not smooth and require 
the apparatus of operator transfer 
matrices to give them a proper interpretation.
Here, we are satisfied keeping $tr[M^\dagger , M] = 0$ even at infinite
$k$ and there are no subtleties in taking the limit. 

The overall scale of $M$ does not matter, so the single parameter we have
is $w\equiv{a\over b}$. Let $e^{(j)}$ be the $j^{\rm th}$ eigenvector
of $M^\dagger M$. 
$$
e^{(j)}={\pmatrix {u_1\cr u_2 \cr \vdots\cr u_k\cr}}^{(j)},~~~~
(J-wN_k ) e^{(j)} =\lambda_j e^{(j)}. \eqno{(3.7)}$$
The equations satisfied by the components $u_i$, with $u_0\equiv 0$
and $u_{k+1} \equiv -w u_k$,
$$
\eqalign
{u_2 = &2\lambda u_1 \cr
u_1+u_3= &2\lambda u_2 \cr
\vdots &\cr
u_{j-1} + u_{j+1} =&2\lambda u_j\cr
\vdots &\cr
u_{k-1}+u_{k+1} =&2\lambda u_k\cr}\eqno{(3.8)}$$
are the recursion relations for the Chebyshev polynomials, and
the initial condition selects the Chebyshev polynomials of second
kind.
$$
u_i \propto U_i (\cos \theta ) (\propto U_i (\cosh \theta ) ) \equiv 
{{\sin (i\theta )}\over {\sin (\theta ) }} \left ( = 
{{\sinh (i\theta )}\over {\sinh (\theta ) }}\right )
\eqno{(3.9)}$$
The argument $\theta$ is determined by 
the $k^{\rm th}$ equation which is equivalent to
$$ U_{k+1}(\lambda) +wU_k (\lambda )=0.\eqno{(3.10)}$$
Let $\lambda_j$ be one of the roots of the above
equation. Then, the eigenvalue
of $(J-wN_k )$ associated with the $(j)$
eigenvector is $\cos (\theta ) =\lambda_j$ ($\cosh (\theta ) =\lambda_j$).

As $\lambda$ varies between $-1$ 
and $1$ the ratio ${{U_{k+1} (\lambda )} \over {U_k (\lambda )}}$ goes 
through $k$ zeros and $k-1$ poles. This implies at least $k-1$ 
solutions for any $w$. The single 
question remains whether a $k^{\rm th}$
solutions fits into the interval. 
This depends on the value of $w$ and the 
values of the ratio at the end points. It is important to
realize that, as long as the Jacobi 
matrix generates orthogonal polynomials with
a positive measure the structure will 
be the same. Our case is particularly simple,
so we can be more explicit, but the structure 
would hold for much more general
mass matrices. Explicitly, we have two cases:
\item {\romannumeral1} $|w|\le 1+{1\over k} $: all roots are in $[-1,1]$. 
\item {\romannumeral2} $|w| > 1+{1\over k} $: 
there are $k-1$ roots in $(-1,1)$
and one root outside this interval. The outside root has the opposite sign
of $w$. For large $k$, due to the rapid 
growth of the $U_k$ polynomials outside
the $[-1,1]$ interval we have, up to exponential corrections (in $k$),
$$
\lambda_k = -{1\over 2} (w +w^{-1}).\eqno{(3.11)}$$

These eigenvalues determine directly the 
eigenvalues of $M^\dagger M$, $m^2_{RL}$.
Thus, for $|{a \over b}| > 1+{1\over k}$ 
we have $k-1$ masses obeying $(a\pm b)^2
\le m^2_{RL} \le (a \mp b)^2$, i.e. typically, they are all
bounded away from zero. There is one mass outside the interval, which, up
to relative corrections exponentially small in $k$ is given by
$$
m^2_{RL}  = {{|b|^{2k}}\over {|a|^{2k+2}}} (a^2 - b^2 )^2,\eqno{(3.12)} $$
If $|{a \over b}| \le 1+{1\over k}$ all $k$ masses are bounded away from
zero by the smaller among $(a\pm b)^2$. 
The above equation 
for $m^2_{RL}$ is derived by setting $|w|=e^{\theta}(1+\delta)$ 
($\theta > 0$) and 
computing $\delta$ to leading order in $|w|^{-k}$. 
In the free case, 
at zero momentum, $b=M_0$ and $a=-1$, so that
$$
m^2_{RL}= |M_0|^{2k} (1-|M_0|^2 )^2 (1+{\cal O}( |M_0|^{2k} )) .
\eqno{(3.13)}$$
The above formula was first derived in [\vranas]. To be precise,
we really only computed at strictly zero momentum, so (3.13) also
contains a wave function renormalization constant. But, from
(2.22) we can infer that the latter has a finite large $k$ limit. 

The theory of orthogonal polynomials 
indicates a certain robustness in the above
structure. Let us focus on $M^\dagger M$. Assume it is a moderate deformation
of the above very simple structure. 
As a first step apply Lanczos tri-diagonalization 
starting from the eigenvector corresponding to the almost massless state
in the unperturbed case. As long as the deformation is small, the positivity
of the measure associated with the 
tri-diagonal matrix (viewed as a Jacobi matrix)
will be maintained. It is important to understand that we think
for the time being about a tri-diagonal matrix for any $k$, in effect
and infinite one, with well defined sequences determining its entries. 
The truncation only enters through the matrix $N_{k}$. Therefore,
the robustness of the matrix $J$ is on the same footing as the robustness of
the exactly massless eigenvalue in the infinite flavor case, where the
index (sometimes called the ``deficit'' in older orthogonal
polynomials literature) is active. But, given this robustness, the effects
of the truncation follow simply from the most basic
properties of any set of orthogonal polynomials.

We conclude therefore that using the theory of orthogonal polynomials
we can convince ourselves that for perturbations under which
the infinite flavor problem is stable there will be
two regimes depending on a parameter $w$: One will have $k-1$ massive
fermions and one very light fermion (for 
finite but large $k$). The complementary regime
will have $k$ heavy fermions. In short, 
the robustness of the infinite flavor
limit implies robustness of the 
hierarchy for large but finite numbers of flavors.

It should be mentioned here that the entrance of orthogonal polynomials
can be understood also as follows: At $k=\infty$ there is an
infinite $M^\dagger M$ and the (positive)
mass square spectrum has a disjoint support
in two components. The first component is 
a $\delta$-function at the origin and the second is a segment supporting
a continuum with typical square root spectral
singularities at both ends of the segment.
If heavy flavor space is continuous in addition to being
infinite the upper bound on the spectrum disappears [\plbfirst]
and the segment gets replaced by a semi-infinite line.
With discrete flavor we can truncate to finite integer $k$ 
and the above spectral
weight gets approximated by a discrete set of abscissas and
associated weights. The $\delta$ function is replaced by one
abscissa and shifts away from the origin by a small amount. 
The truncation
amounts just to the well known Gaussian quadrature we are familiar
with from numerical analysis, 
where the formulae are indeed derived using 
the theory of orthogonal polynomials.

The above discussion about the mass matrix is
somewhat more general than
the one in [\shamir] and makes no reference to specific properties
of the propagator. Nevertheless, it is not more illuminating because
it is a bit mathematical and does not tell us much about not too
large $k$'s, which is, after all, what matters in practice.

To really understand what goes on for moderate $k$ consider the
light mass formula for $k=2$ and $a^2 >> b^2$. 
Every Particle physicist would be immediately
reminded of the see-saw mass formula [\seesaw]. Therefore, there should 
exist a more familiar way to understand the hierarchy and its robustness. 
For this we need one more basis change. 

Let us rotate the right handed fermions alone with the intention to make
$M$ hermitian. Then we can diagonalize $M$ itself, not only $M^\dagger M$
or $MM^\dagger$. A simple relabeling of fields yields:
$$
M=\pmatrix { 0&0&\ldots&0&0&b\cr
	     0&0&\ldots&0&b&a\cr
 	     0&0&\ldots&b&a&0\cr
	     \vdots&\vdots&\vdots&\vdots&\vdots&\vdots\cr
	     b&a&\ldots&0&0&0\cr}.\eqno{(3.14)}$$

Assume now that $|b|<<|a|$. We work to leading order in the ratio. First
set $b=0$. $M^2$ is then diagonal and has $k-1$ eigenvalues equal to 
$|a|^2$ and one eigenvalue equal to zero. The nonzero eigenvalues will
not be affected much by turning on $b$. We know their leading behavior already.
It remains to find out what happens to the lightest mode.
Note that:
$$|\det M | = |b|^{k} .\eqno{(3.15)}$$
Therefore,
$$
m^2_{RL} = {{|b|^{2k}}\over {|a|^{2(k-1)}}}
\eqno{(3.16)}$$
to leading order in ${{|b|}\over {|a|}}$. 
The expression is consistent with the previous computation
at large $k$ and moderate ${{|b|}\over {|a|}}$.
So, if the ratio of the  entries is small, for any
finite $k$ we get an amplification 
of this ratio resulting in a large hierarchy.
This is the see-saw mechanism. It is obvious now that the
above considerations would hold if we replaced 
the constant entries $b$ by varying entries $b_i$,
and similarly the entries $a$ by $a_i$, 
as long as the orders of magnitude stay the same. 
We shall still obtain the same rough hierarchy. 

However, if we add a non-zero entry 
in the $(1,1)$ corner of our hermitian $M$, 
there is a dramatic change, 
the determinant of $M$ going from being very small
to being of order $a^{k-1}$. 
The light state is completely lost. What promises
us that any such entry of $M$, 
when generated radiatively, will be so small
that it will preserve the hierarchy ? Clearly, there must
be some way to understand the orders of magnitude of all entries
of $M$ that can be radiatively generated. Some approximate conservation
law must protect these entries from growing too much
as they move away from the main off-diagonal. 
The mechanism that does that has been invented a long
time ago by Froggatt and 
Nielsen [\fn].\footnote{*}{I thank Y. Kikukawa for reminding
me of this mechanism.} 

The F-N mechanism, can be made to work as follows in our context:
In the original expression for $D$ replace each $B$-entry above the
diagonal by $(y\phi)B$ and each $B$-entry below the diagonal by
$(y\phi)^* B$. $\phi$ is a unit 
charged scalar Higgs field. The unit charge is
with respect to a new abelian group which is spontaneously broken.
The fermions are also charged under this group; we denote this new
charge by $F_N$. We pick an assignment of charges for the fermions
as defined in figure 1 for $k=5$.

\figure{4}{\captionfour}{5.3}

The intuitive picture is best described by first simplifying
and ignoring all doublers: 
For some reasonably weak $y$ (not fine tuned, but small)
in the presence of spontaneous $F_N$ symmetry breakdown, the
lattice fermion spectrum can be viewed, in continuum language,
as follows: There are $k-1$ massive Dirac fermions,
all having masses of the same order (the condensate is of the
order of the lattice cutoff). All these Dirac fermions
realize the $F_N$ $U(1)$ symmetry vectorially. 
There also is one light Dirac
fermion whose left and right components carry 
very different $F_N$ charges, thus realizing 
the $U(1)$ symmetry chirally.
These two components can only communicate, for perturbative $y$,
via $k$ ``exchanges'' with the condensate. All intermediate fermions
in the diagram are massive. If the ratio between the 
condensate scale $<\phi>$ and the heavy mass is 
sufficiently small relative to 
unity (but not necessarily unnaturally so)
we get our hierarchy. 

But, we shouldn't ignore the doublers. 
They cannot be ignored because the $F_N$ $U(1)$ is an {\sl exact symmetry}
on the lattice, and therefore, in a continuum limit that
{\sl also maintains the Higgs field}, it must be realized in a vectorial way
to avoid anomalies. On the other hand, as long as we do not
keep the Higgs field in the continuum limit and are in the
spontaneously broken phase, the intuitive picture would lead
to a selfconsistent perturbation theory.
All we really want 
from this picture is an intuitive reason
for why, when we calculate, we expect to get a mass hierarchy,
and why radiative corrections should not be suspected to immediately
destroy it. 

More formally, we can eliminate the Higgs field and replace it by
a constant and just observe that
making an $F_N$ global
gauge transformation on all the 
fermions with parameter $\chi$ and, simultaneously, 
changing $y$ by the phase $e^{i\chi}$ leaves
the theory invariant. Therefore, the two point function involving the
left and right components of the quark at small
momenta which would become massless if $y$ were set to zero, must 
have a dependence on $y$ of the 
form $y^k f_k (|y|^2;{\cal A})$. ${\cal A}$ 
represents the gauge background. 
By a finite rescaling of $y$ we can arrange 
for $f_k (0;{\cal A})$ to be of order unity. 
If the rescaled $y$ is only 
reasonably small (say of order .5), an exponential mass hierarchy 
gets generated for large $k$'s. 

There is one problem we have not directly addressed yet, but we must. 
The above
arguments indicate that the light quark will stay light when
gauge interactions are turned on. They also might be taken,
in particular in the domain wall picture [\fursham], to indicate
that the left and right components of the light quark completely
decouple. If this is true they can be rotated independently,
and then we have an inescapable $U(1)_{L-R}$ problem! Since
we know that this global symmetry is explicitly broken, the
arguments cannot be completely correct, and if they aren't,
how can we trust them to correctly indicate approximate masslessness?

The answer is known at the mathematical level: In the overlap
the $U(1)$ problem is solved in exactly the way discovered
by 't Hooft. The independent rotations on the decoupled
components of the massless quark are not really rigorously
defined in the path integral since $k$ is strictly infinite.
This creates a loophole just where needed in order to make
't Hooft's solution to
the axial $U(1)$ problem work on the lattice.  
Since the truncated model approaches the overlap, 
the mechanism preserving masslessness has to allow $U(1)$ breaking
just as needed. Moreover, from our exact expression for the fermion
determinant in the presence of the $\mu$-mass term we see
explicitly that gauge configurations carrying nontrivial
topology make large contributions. For example, for a single flavor,
an instanton makes an unsuppressed contribution 
to $<(\bar\psi \psi )_{\rm phys}>$.

In the intuitive picture based on the
F-N mechanism, the role of nontrivial topology of the
gauge background can be understood as follows:
Suppose we have several copies of the whole  setup, with an exact
vectorial $SU(N_f)$ symmetry. 
In the intuitive picture we described
before the $F_N$ $U(1)$, being chiral, is also explicitly
broken by instantons and the $F_N$ charge is not fully conserved.
This lack of conservation makes it possible for a correlation
function containing $2N_f$ chiral components of the light quarks
to become unprotected, and not necessarily small. 
Of course, on the lattice this picture is not really correct
since the $F_N$ symmetry is not chiral.
The single correct statement we can make is that we expect those
functions $f_k (0;{\cal A})$ (or their appropriate generalizations
for expectation values of products of more
than two fermion fields) that get instanton contributions to
be enhanced and avoid the suppression that generates
the hierarchy. 

In some loose sense, it seems that the Smit-Swift [\smitswift]
approach
to use Yukawa couplings for obtaining chiral symmetries
in the continuum might work for almost massless 
vector-like theories once
it is combined with the F-N mechanism. The Yukawa couplings
indeed can remove the doublers while keeping the ``desired''
fermion relatively light. The contribution 
of the F-N mechanism is to take a 
mass ratio between the doubler and the light fermion
of order $2$ say and amplify it to $2^k$. Since the heavy
fermions have a mass of order 1 in lattice units, one gets
a very light fermion. Thus, what Yukawa models were
unable to do, even with fine tuning, namely generate
some large mass hierarchies between charged (under the group
that is gauged) fermions and other fermions, is no longer
needed. Only some ``start'' in the right direction is needed
and then F-N's mechanism can amplify the mass hierarchy
as much as one wants. If one wishes strictly massless quarks
one would need infinite amplification, so we are back in the overlap
case. Therefore, if we wish to regulate a chiral gauge theory
we cannot avoid an infinite number of fermions. But, in the
vector like case, where very small quark masses are sufficient,
the combination of the Smit-Swift and F-N mechanisms seems
useful. 
So, despite some
failures for chiral gauge theories [\jansen] (even when combined 
with truncated fifth dimension models), a re-examination
of Yukawa models, this time with the intention to provide
a cleaner numerical approach to almost massless QCD, might 
produce something new and useful. One may also speculate that 
integrating out the heavy F-N Higgs fields, which would leave
behind some multilinear fermion interactions, could provide a 
connection to the approach taken recently 
in [\kogsinc], also with the purpose to get closer to the
chiral limit at finite lattice spacings. It would be nice
to unify the different tricks 
people have come up with while attempting
to get global chiral symmetries respected by the lattice.

Very recently [\taniguchi] perturbative calculations in the gauge coupling
constant have been undertaken to show robustness. Full details
are not available at the moment, but the conclusion is that
the hierarchy is maintained. This is expected, although the
explicit check is definitely reassuring. But, we should keep
in mind that perturbative checks 
of this kind will
not be able to address the $U(1)$ problem mentioned above,
and therefore, an intuitive picture should be welcome.

\vskip .5in
\centerline{\bf 4. Summary and Outlook}

In this paper we derived some exact expressions for the integrals
over lattice fermions in systems that regularize 
vector like continuum theories containing very light fermions of Dirac
or Majorana type. These expressions make explicit the effect
of the truncation involved in going from the overlap where
the fermions are strictly massless to actions that can be 
simulated numerically with relative ease. In addition to the
Dirac case we also dealt with the Majorana fermions needed
to simulate pure gauge supersymmetric theories. A criterion
for the goodness of the truncation approximation was given in
terms of certain extremal eigenvalues of the
transfer matrix. Potential difficulties related to topology changes
were noted. 

We also presented several different views that explain the stability
of the large mass hierarchy between the lattice cutoff and the
fermion mass. We found that two well known mechanisms, the see-saw
and the Froggatt-Nielsen mechanism are alternatives to the
extra dimension view. This understanding should contribute to
our ability to trust and correctly interpret numerical results 
obtained using the truncated overlap. 

Considering the application of 
Renormalization Group ideas to the truncated overlap, we suggest 
that the blocking transformation should also integrate out 
some of the $k$ heavy fermions.\footnote{*}{Specifically: we
already know from [\balaban] that using $\delta$-function constraints
in the fermion path integral to implement 
the thinning of degrees of freedom is a bad idea, but if one
uses smooth kernels for the same purpose no undesired
non-localities get generated. We now imagine a kernel that
not only implements the usual thinning but simultaneously
reduces the flavor
number $k$.} Indeed, the pure glue correlation length 
increases exponentially with the gauge coupling $\beta$ while the
free fermion correlation length increases exponentially with $k$.
In other words the thinning of degrees of freedom seems to be naturally
applicable to both real and flavor spaces simultaneously. 
A flow in the $k-\beta$ space could determine the anomalous dimension
of the quark mass. It would be interesting to investigate this 
further.  

One way to translate our insights into the structure of the mass matrix into
something practical is to devise ways to accelerate further the
speed of hierarchy generation as a function of $k$. In four
dimensions it appears that one can obtain useful numbers 
with $k\sim 10-15$ ([\blumsoni]) while in two dimensions 
somewhat lower $k$'s might be adequate, although this is not
completely clear ([\jaster],[\vranas]). 
The small effective mass goes essentially
as $(\kappa(\beta ))^k$ where $0<\kappa (\beta ) < 1$ is some 
unknown function. Any trick that reduces $\kappa (\beta )$ would
allow a reduction of $k$ without changing the effective mass. 
At $\beta=\infty$ $\kappa (\beta )$ is governed by the ratio of
the $B$ $L-R$ term to the $-1$ $L-R$ term in the matrix $D$. 
We can make the ratio as small as we want at zero momentum,
but there is an increase as we move away from the origin.
A simple generalization of the action that could ameliorate the
effect, and thus, hopefully, end up decreasing the required $k$,
is as follows:

Find two polynomials $p(\lambda )$ and $q(\lambda )$ which have
the following properties:
\item{1.} For $\lambda \in [0,2d]$, $p(\lambda ) , q(\lambda ) > 0$.
\item{2.} There exists a real number $0<r<2$ such that, 
for $\lambda \in (0,r)$, $0<{{p(\lambda )}\over {q(\lambda )}} <<1$ while,
for $\lambda \in ( 2,2d)$,  $0<{{p(\lambda )}\over {q(\lambda )}} \ge 1$.

Then replace the entries $B$ in $D$ by $p(B_0 )$ and the entries
$-1$ by $-q( B_0 )$. The truncated model we analyzed in this paper
had $p(\lambda ) = \lambda + M_0$ and $q(\lambda) =1$. In
addition to being small we would also like the 
ratio $0<{{p(\lambda )}\over {q(\lambda )}}$ to be relatively 
constant in $(0,r)$ while it undergoes a rapid increase in $[r,2]$
to some value slightly higher than unity and stays above unity
(but not necessarily with small variability) in the 
entire interval $[2,2d]$. Increasing the degrees of the polynomials
carries with it some computational cost, but, it seems that this
cost could be compensated by needing a smaller $k$-value. We hope
to return to this issue in future work. 

Another way to put the viewpoints of this paper to some use
is to reinterpret some of the data obtained by Blum and Soni
[\blumsoni]
in the quenched approximation. 
They obtained a few values for the pion mass square at
different $\beta$'s, various $k$'s, with values of $M_0$ chosen
so that one works in the triangular area of figure 2.
For their largest $\beta$ value, they found that they needed
to increase $k$ from $10$ to $14$ in order to obtain apparent
massless quarks in the limit $\mu =0$ (masslessness was
determined by extrapolation
from finite $\mu$ values). Since we view the system at 
finite $k$ as having a quark mass coming from two sources,
one the finiteness of $k$ and the other $\mu$, the data 
at $k=10$ extrapolated to $\mu =0$ can be taken as an
indirect measurement of the ``F-N suppressed'' quark mass
at $k=10$. In other words, the finite small 
pion mass is proportional to the sum of the 
truncation-induced quark mass and the explicit quark mass term. 
A test at a few smaller values of $k$ might have
provided numerical evidence for the exponential dependence
on $k$. On the other hand it was found, 
at the highest $\beta$ value, that the system behaved
as if the  truncation induced quark mass was negative.
This was ascribed to a numerical detectable higher order
term in the chiral effective Lagrangian (possibly with quenching
effects included). A simpler possibility is that the
truncation induced quark mass is indeed negative. We expect
that when one works at parameters corresponding to the
interior of the triangle of figure 2 to the right of the 
y-axis one induces a positive quark mass, but what happens
more to the left is not completely understood. It is possible
that, at $\beta=\infty$ the values $-1 < M_0 <0 $ induce, at
finite $k$, finite negative quark masses (in the sense that
includes the QCD $\theta$ parameter). 

The last conjecture may be a bit mystifying given the overt
positivity of the regularized fermion determinant in (2.12) for example.
So, let us clarify the issue: The basic question is whether the matrix
$B$ can also have negative eigenvalues for some gauge backgrounds. 
As long as $B$ is positive definite we are sure that the
fermion determinant is always positive, including for backgrounds
containing a single instanton. Therefore, the quark mass must be
positive. If $B$ has a negative eigenvalue the basic 
formula $T=e^{-H}$, ostensibly defining a hermitian $H$, breaks down.
Indeed the expression for $T$, which (unlike in the transfer
matrix derivation of [\npold] where $\sqrt {B}$ 
was used at intermediary steps) is now  valid as long as $\det B \ne 0$
shows that $T$ is positive definite iff $B$ is such:
$$
T=\pmatrix { 0&1\cr -1 & C^\dagger\cr } \pmatrix {B & 0 \cr 0 & {1\over B}\cr}
\pmatrix { 0& -1\cr 1& C\cr }\eqno{(4.1)}$$
All formulae containing $H$ can be written in terms of $T$ and are
valid as long as $\det B \ne 0$. With a $T$ that can have 
negative eigenvalues the fermionic determinants do not have to be
always positive any more. On the other hand, the structure
of the mass matrix still indicates the presence of very light
quarks. Therefore, we conjecture that to the left of the y-axis
we are dealing with light quarks but with a negative mass.
It is hoped that this
point would be clarified in future work.

Before closing, let us mention that the insights of this paper
are hoped (as mentioned in [\recent])
to be of some help also to attempts to regularize chiral
gauge theories.

\vskip .2in

\noindent {\bf Acknowledgments}:
This research was supported in part by the 
DOE under grant  \# DE-FG05-96ER40559. I am grateful to Yoshio
Kikukawa for his comments in the context of the Froggatt Nielsen
mechanism, his detection of several errors in an earlier draft 
and many discussions on the topics of this paper. I also wish
to acknowledge e-mail comments by  T.-W. Chiu, I. Montvay and  S. Zenkin.
I am also indebted to Y. Kikukawa and P. Vranas for comments on 
a very recent draft of this paper.

\vskip .3in

\centerline{\bf Appendix}
The purpose of this appendix is to derive a general formula
for the determinant of a block tri-diagonal matrix with non-vanishing
corner blocks. This formula is analogous to one employed
by Gibbs and followers in the numerical
study of finite density effects in QCD [\gibbs].
Let the matrix $D$ have even dimension $n=2k$
where each entry is a $q\times q$ block:
$$
D=\pmatrix{B_1 & A_1 & 0& 0 &\ldots & 0 & 0& C_n\cr
	   C_1 & B_2 & A_2& 0&\ldots & 0 & 0 & 0\cr
	   0&  C_2 & B_3 & A_3&\ldots & 0 & 0 & 0\cr
	   0&0  &C_3 & B_4 &\ldots & 0 & 0 & 0\cr
	   \vdots&\vdots&\vdots&\ddots&\ddots&
	   \ddots &\vdots&\vdots\cr
	   \vdots&\vdots&\vdots&\vdots&\ddots&
	   \ddots &\ddots&\vdots\cr
	   \vdots&\vdots&\vdots&\vdots&\vdots&
	   \ddots &\ddots&\ddots\cr
           A_n&0&0&0&\ldots&0&C_{n-1} &B_n\cr}.\eqno{(A.1)}
$$ 
Moving the leftmost column of blocks into the place of the rightmost
column we obtain the matrix $D^\prime$:
   $$
D^\prime
=\pmatrix{A_1 & 0& 0 &\ldots & 0 & 0& C_n& B_1 \cr
	  B_2 & A_2& 0&\ldots & 0 & 0 & 0 & C_1 \cr
	    C_2 & B_3 & A_3&\ldots & 0 & 0 & 0&0 \cr
	   0  &C_3 & B_4 &\ldots & 0 & 0 & 0&0\cr
	   \vdots&\vdots&\ddots&\ddots&
	   \ddots &\vdots&\vdots&\vdots\cr
	   \vdots&\vdots&\vdots&\ddots&
	   \ddots &\ddots&\vdots&\vdots\cr
	   \vdots&\vdots&\vdots&\vdots&
	   \ddots &\ddots&\ddots&\vdots\cr
          0&0&0&\ldots&0&C_{n-1} &B_n &A_n\cr}.\eqno{(A.2)}
$$  
Counting minus signs we obtain, using the evenness of $n$:
$$\det D = (-)^q\det D^\prime . \eqno{(A.3)}$$

We now view $D^\prime$ as a $k\times k$ matrix  
with $2q\times 2q$ blocks, defined below:
$$\eqalign{
\alpha_1 =& \pmatrix { A_1 & 0 \cr B_2 & A_2 \cr },
\alpha_2 = \pmatrix { A_3 & 0 \cr B_4 & A_4 \cr },\dots ,
\alpha_j = \pmatrix { A_{2j-1} & 0 \cr B_{2j} & A_{2j}\cr },
\dots ,
\alpha_k = \pmatrix { A_{n-1} & 0 \cr B_n & A_n \cr }\cr
\beta_1 =& \pmatrix { C_2 & B_3 \cr 0 & C_3 \cr },
\beta_2 = \pmatrix { C_4 & B_5 \cr 0 & C_5 \cr },\dots ,
\beta_j = \pmatrix { C_{2j} & B_{2j+1} \cr 0 & C_{2j+1} \cr },
\dots ,
\beta_k = \pmatrix { C_n & B_1 \cr 0 & C_1 \cr }\cr}
\eqno{(A.4)}$$
In terms of the $\alpha$ and $\beta$ blocks we have
$$
D^\prime=
\pmatrix{\alpha_1&0&0&\ldots&0&\beta_k\cr
          \beta_1&\alpha_2&0&\ldots&0&0\cr
	  0&\beta_2 &\alpha_3&\ldots&0&0\cr
	  \vdots &\vdots &\ddots &\ddots&\vdots &\vdots\cr
	  \vdots &\vdots &\vdots &\ddots &\ddots&\vdots \cr
	  0&0&0&\ldots&\beta_{k-1}&\alpha_k\cr}.\eqno{(A.5)}$$
We now consider the following linear equations for the unknown
$2q\times 2q$ matrices $v_j, j=1,2,\dots,k$:
$$
D^\prime=
\pmatrix{\alpha_1&0&0&\ldots&0&0\cr
          \beta_1&\alpha_2&0&\ldots&0&0\cr
	  0&\beta_2 &\alpha_3&\ldots&0&0\cr
	  \vdots &\vdots &\ddots &\ddots&\vdots &\vdots\cr
	  \vdots &\vdots &\vdots &\ddots &\ddots&\vdots \cr
	  0&0&0&\ldots&\beta_{k-1}&\alpha_k\cr}
\pmatrix{1&0&0&\ldots&0&-v_1\cr
          0&1&0&\ldots&0&-v_2\cr
	  0&0 &1 &\ldots&0&-v_3\cr
	  \vdots &\vdots &\ddots &\ddots&\vdots &\vdots\cr
	  \vdots &\vdots &\vdots &\ddots &\ddots&\vdots \cr
	  0&0&0&\ldots&0&1-v_k\cr}.\eqno{(A.6)}$$

More explicitly, the equations for the $v_j$ are:
$$
\eqalign{ ~\alpha_1 ~v_1 ~+ &~~~~~\beta_k =
0\cr  ~\alpha_2 ~v_2 ~+ &~~~\beta_1 ~v_1 =0\cr
	  \dots&\dots\cr
	  \alpha_k ~v_k ~+&~\beta_{k-1}~ v_{k-1}=0\cr}\eqno{(A.7)}$$
With $v_j$ solving the above equations, we obtain:
$$
\det D^\prime = [\prod_{j=1}^{k-1} \det \alpha_j ] 
\det (\alpha_k - \alpha _k v_k ).\eqno{(A.8)}$$
The single unknown is  $v_k$, and the solution is given below:
$$
\alpha_k v_k = (-)^k \beta_{k-1} \alpha_{k-1}^{-1} 
\beta_{k-2}\alpha_{k-2}^{-1} \dots \beta_2 
\alpha_2^{-1}\beta_1 \alpha_1^{-1} \beta_k .\eqno{(A.9)}$$

We obtain, finally, 
$$
\det D = (-)^q ~[\prod_{j=1}^{k-1}\det \alpha_j ] 
\det [\alpha_k + (-\beta_{k-1} \alpha_{k-1}^{-1} )
(-\beta_{k-2} \alpha_{k-2}^{-1} )\dots 
(-\beta_{2} \alpha_{2}^{-1} )(-\beta_{1} \alpha_{1}^{-1} )\beta_k ].
\eqno{(A.10)}$$
We chose a form well suited to our applications. The above
form can be used to derive propagators by varying with respect to
various entries of $D$.

Our main applications have:
$$
\eqalign{
\alpha_j =& \pmatrix{B&0\cr -C&-1\cr} 
\equiv \alpha,~~~j=1,\dots,k-1~~\alpha_k =\pmatrix {B&0\cr -C& X\cr}
\cr
\beta_j =&\pmatrix {-1& C^\dagger\cr 0&B\cr}\equiv\beta
,~~~j=1,\dots,k-1~~\beta_k =
\pmatrix {Y&C^\dagger\cr 0& B}
\cr}
\eqno{(A.11)}$$
We shall also need
$$
\alpha^{-1} =\pmatrix {{1\over B} &0\cr -C {1\over B} & -1\cr}.
\eqno{(A.12)}$$
In this case we obtain:
$$
\det D = (-)^q (\det B)^{k-1} (-)^{q(k-1)} \det [\alpha_k -
\alpha (-\alpha^{-1} \beta )^k \beta^{-1} \beta_k ].\eqno{(A.13)}$$
The product $-\alpha^{-1} \beta $ is related to the transfer matrix $T$:
$$
-\alpha^{-1} \beta = \pmatrix { {1\over B} & -{1\over B} C^\dagger \cr
-C{1\over B}& C{1\over B}C^{\dagger} +B\cr} \equiv 
\pmatrix {0&1\cr 1&0} T^{-1} \pmatrix {0&1\cr 1&0}$$
$$
T=\pmatrix {{1\over B} & {1\over B} C \cr
         C^\dagger {1\over B} & C^\dagger {1\over B} C +B \cr}.
	 \eqno{(A.14)}
$$
Finally, we can write:
$$
\det D = (-)^{q(k-1)} (\det B)^k \det 
\left [ \pmatrix {-X &0 \cr 0& 1 } - T^{-k} \pmatrix {1&0\cr 0 &-Y\cr}
\right ].\eqno{(A.15)}$$

\vskip .3in

\noindent{\bf References}
\vskip .1in
\item {[\recent]} H. Neuberger, hep-lat/9707022.
\item {[\plbfirst]} R. Narayanan, H. Neuberger, Phys. Lett. B302 (1993) 62.
\item {[\npold]} R. Narayanan, H. Neuberger, Nucl. Phys. B. 412 (1994) 574.
\item {[\npblong]} R. Narayanan, H. Neuberger, Nucl. Phys. B443 (1995) 305.
\item {[\blumsoni]} T. Blum, A. Soni, Phys. Rev. D56 (1997) 174, 
hep-lat/9706023.
\item {[\jaster]} A. Jaster, hep-lat/9605011.
\item {[\vranas]} P. Vranas, hep-lat/9705023, hep-lat/9709119.
\item {[\kaplan]} D. B. Kaplan, Phys. Lett. B288 (1992) 342.
\item {[\shamir]} Y. Shamir, Nucl. Phys. B406 (1993) 90.
\item {[\fursham]} V. Furman, Y. Shamir, Nucl. Phys. B439 (1995) 54.
\item {[\nninfmass]} R. Narayanan, H. Neuberger, 
Phys. Lett. B393 (1997) 360.
\item {[\kiknn]} Y. Kikukawa,  R. Narayanan, H. Neuberger, 
Phys. Lett. B399 (1997) 105.
\item {[\narvra]} R. Narayanan, P. Vranas, hep-lat/9702005.
\item {[\boya]} D. Boyanovsky, E. Dagotto, 
E. Fradkin, Nucl. Phys. B285 (1987) 340.
\item {[\creutzbnd]} M. Creutz, I. Horvath, Phys. Rev. D50 (1994) 2297.
\item {[\shotkey]} W. Shockley, Phys. Rev. 56 (1939) 317; 
W. G. Pollard, Phys. Rev. 56 (1939) 324.
\item {[\massineq]} S. Nussinov, 
Phys. Rev. Lett. 51 (1983) 2081, 52 (1984) 966; 
D. Weingarten, Phys. Rev. Lett. 51 (1983) 1830; 
E. Witten, Phys. Rev. Lett. 51 (1983) 2351.
\item {[\rebbi]} C. Rebbi, Phys. Lett. B186 (1987) 200.
\item {[\ginspwil]} P. Ginsparg, K. Wilson, Phys. Rev. D25 (1982) 2649.
\item {[\curciv]} G. Curci, G. Veneziano, Nucl. Phys. B292 (1987) 555.
\item {[\aokizen]} S. Aoki, K. Nagai, S. V. Zenkin, hep-lat/9705001.
\item {[\huetnn]} P. Huet, R. Narayanan, 
H. Neuberger, Phys. Lett. B380 (1996) 291. 
\item {[\hsu]} S. Hsu, hep-th/9704149.
\item {[\odddim]} Y. Kikukawa, H. Neuberger, hep-lat/9707016.
\item {[\montvay]} G. Koutsoumbas, I. Montvay, Phys. Lett. B398 (1997) 130; 
\item {} I. Montvay, hep-lat/97090080.
\item {[\nishimura]} J. Nishimura, hep-lat/9709112; 
Phys. Lett. B406 (1997) 215;
T. Hotta, T. Izubuchi, J. Nishimura, hep-lat/9709075.
\item {[\italiansusy]} A. Donini, M. Guagnelli, 
P. Hernandez, A. Vladikas, hep-lat/9710065.
\item {[\eyalgomez]} E. Cohen, C. Gomez, Phys. Rev. Lett. 52 (1984) 237.
\item {[\seesaw]} M. Gell-Mann, P. Ramond, R. Slansky 
in {\it Supergravity} edited by P. van Nieuwenhuizen 
and D. Z. Friedman (North-Holland, 1979).
\item {[\fn]} C. D. Froggatt,  H. B. Nielsen, Nucl. Phys. B147 (1979) 277.
\item {[\smitswift]} J. Smit, Acta Physica Polonica B17 (1986) 531. 
P.D.V. Swift, Phys. Lett. B145 (1984) 256.
\item {[\jansen]} K. Jansen, Phys. Rep. 273 (1995) 147.
\item {[\kogsinc]} J. B. Kogut, D. K. Sinclair, hep-lat/9607083.
\item {[\taniguchi]} S. Aoki, Y. Taniguchi, hep-lat/9709123.
\item {[\balaban]} T. Balaban, Lett. in Math. Phys. 17 (1989) 209.
\item {[\gibbs]} P. Gibbs, Phys. Lett B172 (1986) 53.

\vfill\eject
\end